\documentclass{ragtime} 
                        
\pdfoutput=1

\newcommand{\M} {\rm\, M}

\title[Perturbing the accretion flow by a passing star]%
      {Perturbing the accretion flow onto a supermassive black hole by a passing star}

\author[P. Sukov\'{a}  et al.
        ]
       {Petra Sukov\'{a}\at{1,a} 
        Michal Zaja\v{c}ek\at{2}
        Vojt\v{e}ch Witzany\at[]{3} 
       \splitauthors
        and Vladim\'{\i}r Karas\at[]{1}\\
        \ins{1}Astronomical Institute of the Czech Academy of Sciences, \splitins[1] Bo\v{c}n\'{\i} II 1401, 141 00, Prague, Czech Republic
        \\
        \ins{2}Center for Theoretical Physics, Polish Academy of Sciences,\splitins[1] Al. Lotnikow 32/46, 02-668 Warsaw, Poland\\
        \ins{3}School of Mathematics and Statistics, University College Dublin,\splitins[1] Belfield, Dublin 4, D04 V1W8, Ireland\\
        \ins{a}\Email{petra.sukova@asu.cas.cz}} 

\graphicspath{{suk/}}

\begin{document}

\begin{abstract}
The close neighbourhood of a supermassive black hole contains not only accreting gas and dust, but also stellar-sized objects like stars, stellar-mass black holes, neutron stars, and dust-enshrouded objects that altogether form a dense nuclear star-cluster. These objects interact with the accreting medium and they perturb the otherwise quasi-stationary configuration of the accretion flow. We investigate how the passages of a star can influence the black hole gaseous environment with GRMHD 2D and 3D simulations. We focus on the changes in the accretion rate and the associated emergence of outflowing blobs of plasma.
\end{abstract}


\section{Introduction}\label{intro}

In the supermassive black hole environment we can expect and in the case of our Galactic center even observationally resolve the presence of stars that form a dense nuclear star-cluster \citep[ and references theirein]{2020ApJ...899...50P}. We can only deduce the number of neutron stars, stellar-mass black holes and other stellar-sized objects originating as an inevitable outcome of the stellar evolution and the feedback processes  \citep{2020A&ARv..28....4N}. If they are indeed embedded in the accretion flow, the mutual interaction between those objects and gas can lead to observable effects, in particular the changes in accretion rate, ejection of plasma blobs and the redistribution of accreting gas. This should then lead to temporal changes of the outgoing radiation.

In the present work we assume that the averaged accretion flow is centered on the plane perpendicular to the rotation axis of the central black hole (the equatorial plane). The field of the black hole is described by the Kerr metric \citep{1973grav.book.....M}. While the gravitational field of the black hole obeys the conditions of axial symmetry and stationarity, the accretion flow can be highly turbulent and non-stationary \citep{2008bhad.book.....K}. 

The motion of stars and the resulting impact on the accretion flow can reveal signatures of the orbital period at the corresponding radius \citep{1994ApJ...422..208K,2016MNRAS.457.1145P}. Therefore, we explore in our contribution the impact of the passages of stars through the accreting medium. We want to understand the effects that this may have on the accretion rate, and we explore whether a fraction of the material can be set on escaping trajectories.

\section{Set-up of the numerical procedure}\label{}
We perform global general-relativistic magneto-hydrodynamical (GRMHD) simulation within the assumed fixed spacetime metric. We compare the results obtained in 2D and 3D simulations of the flow using the publicly available code {\tt HARMPI} \citep{2015MNRAS.454.1848R,2007MNRAS.379..469T}. The adopted numerical tool is based on the original HARM code \citep{0004-637X-589-1-444,2006ApJ...641..626N}, which we have modified in order to explore the effects of mutual interactions between the gaseous medium and the transiting body of a star. The code uses a conservative, shock-capturing scheme with a staggered magnetic field representation and adaptive time step $\Delta t$.
 
 We follow the evolution of gas under the assumption of vanishing resistivity and the polytropic equation of state $p = K \rho^\gamma$ with the adiabatic index $\gamma=13/9$. The background spacetime is described by the Kerr metric with the spin parameter $a=0.5$ (for definiteness of the example), using the modified Kerr-Schild coordinates penetrating below the horizon \citep{1973grav.book.....M}. We defined the grid with logarithmic spacing in the $r$-direction in such a way that there are always at least 5 cells below the horizon. In the $\theta$ direction, the grid is concentrated along the equatorial plane \citep{Sasha-smart-grid}. Thanks to the non-uniform spacing of the grid, we have higher resolution in the region of interest and we do not need to employ the mesh refinement. The outer boundary of the grid is set at $R_{\rm out}=2\cdot10^4\M$. For more details of the numerical setup, see the forthcoming paper (Sukov\'a et al., 2021, work in progress).

\section{Results}\label{results}
We present preliminary results from several computational runs that complement a more detailed discussion in (Sukov\'a et al., 2021, work in progress). In the latter paper we explore the role of the shape and orientation of the stellar orbits, while here we focus more on the effects of the resolution of the computational grid and the exact realisation of the body moving through the gaseous medium. Hence, we will be able to better constrain the possible uncertainties in our results.

\subsection{Effects of the grid resolution}\label{Sec:resolution}
The resolution of the grid influences our simulations in two ways. First, the resolution is crucial to capture the MRI in the flow, thus the accretion rate and the complexity of the flow are affected. Second, the exact description of the star, its shape and minimal possible diameter are constrained by the resolution. Because we let the torus evolve before turning on the perturbation, the state of the flow at the moment when the star starts to orbit the black hole is not the same as the initial conditions. Therefore we first explore the effect of the resolution on the non-perturbed evolution, and we turn to the study of how the action of the star depends on the resolution afterwards. 

\begin{figure}
\begin{center}
\includegraphics[width=\linewidth]{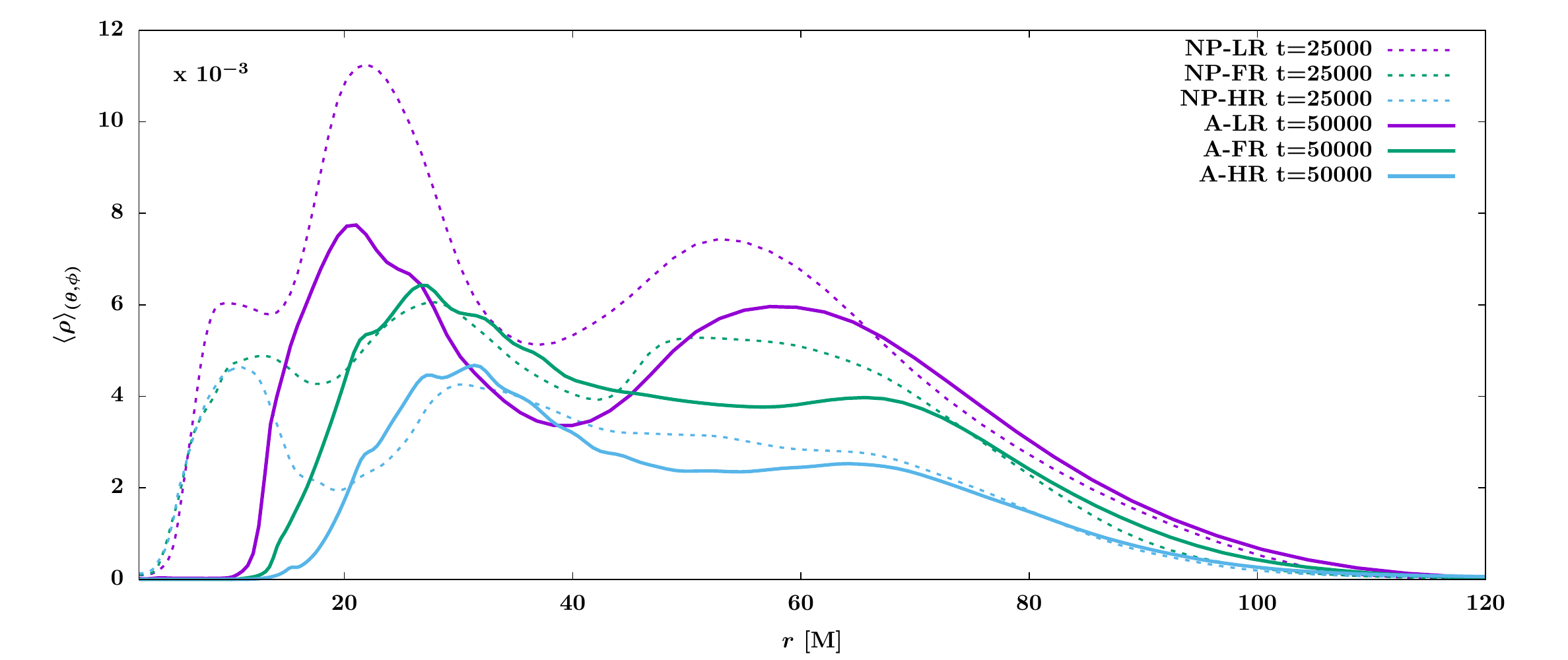}

\includegraphics[width=\linewidth]{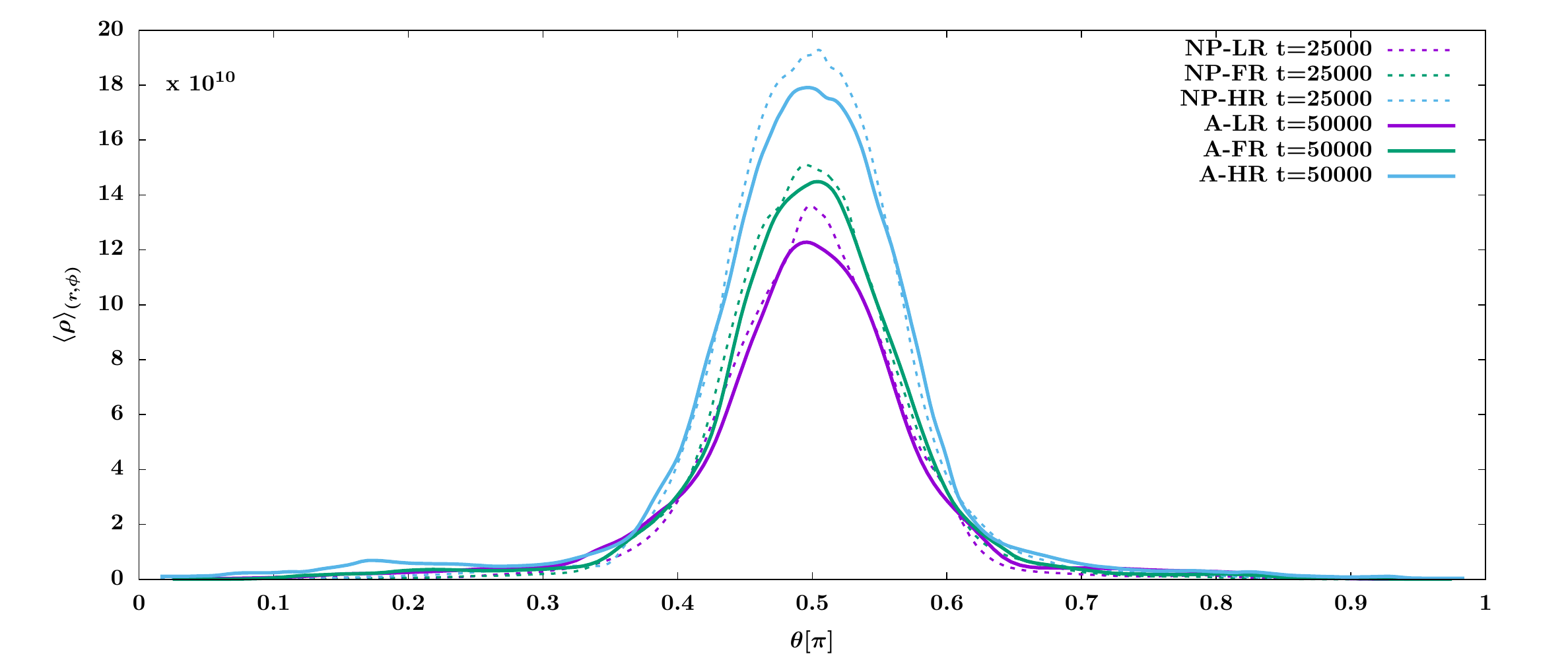}
\end{center}
\par\vspace{-1ex}\par
\caption{\label{Fig:rho}The radial (top) and angular (bottom) profiles of averaged density $\left<\rho\right>_{(\theta,\phi)}(r)$ and $\left<\rho\right>_{(r,\phi)}(\theta)$. The dashed lines show the profile at $t_{\rm in}=25000\M$ for LR (purple), FR (green) and HR (blue) runs; the solid lines are computed at $t_{\rm f}=50000\M$. }
\end{figure}

\begin{figure}
\begin{center}
\includegraphics[width=\linewidth]{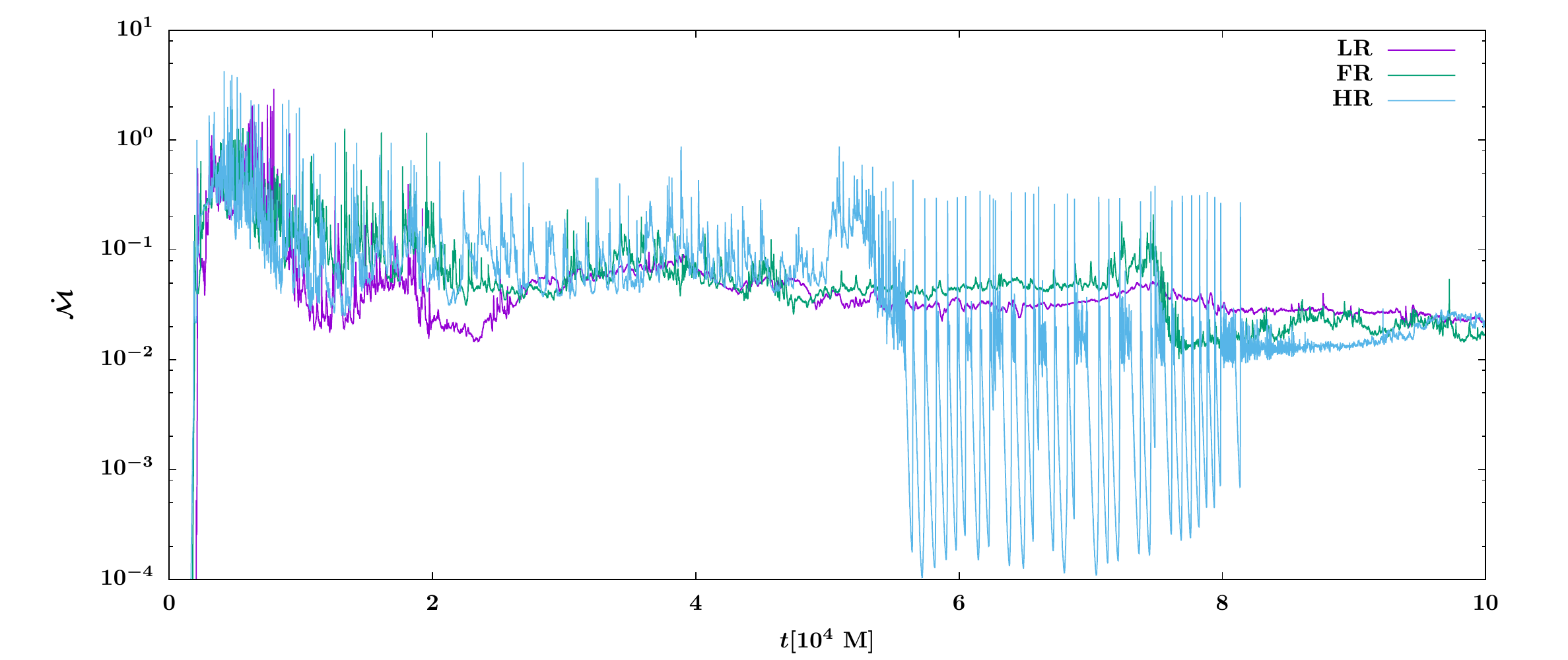}
\end{center}
\par\vspace{-1ex}\par
\caption{\label{Fig:mdot}Time dependence of accretion rate $\dot{\mathcal{M}}$ is plotted for the non-perturbed runs with LR (purple), FR (green) and HR (blue). Frequent, intermittent fluctuations are characteristic.}
\end{figure}

\subsubsection{Non-perturbed evolution of the accretion flow}

Additionally to the 2D runs presented in (Sukov\'a et al., 2021, work in progress), which were computed with the fiducial resolution (FR) of $n_r=252, n_\theta=192$, we present here also the results of runs in the low resolution (LR) of $n_r=192, n_\theta=144$, and the high resolution (HR) set-up of $n_r=384, n_\theta=288$. We initialized the computation with the same parameters for each resolution, which is the torus from the family of solutions introduced by \citet{Witzany_Jefremov-tori} with $\kappa = 7.61,l_0 = 8.46098\M$ stretching from $r_{\rm min}=20\M$ to $r_{\rm max}=90\M$. The torus is threaded by a poloidal magnetic field with field lines that follow the isocontours of density; the gas to magnetic pressure ratio equals to $\beta=p_g/p_m = 100$. The star is described as a gradual perturbation (GP) contained within the full star volume (see Section~\ref{Sec:Geodesic-star} for further details).

We check the ability to describe the MRI in our computations in the following way. At the radius of the density maximum in the initial state we find how many cells $N_\theta$ in $\theta$ direction are used to cover the density scale height $H$, which is defined as the height at which the density decreases to $\rho_{\rm max}/e$. This number turns out to be $N_\theta = 20$ for LR, $N_\theta = 28$ for FR and $N_\theta = 42$ for HR. Then, according to \citet{Hawley_2011} the quality $Q_z$ of the resolution of the unstable MRI modes is estimated as
\begin{equation}
    Q_z \eqsim 0.6 \, N_\theta \left(\frac{100}{\beta}\right)^{1/2}\left(\frac{\langle v_{\mathrm{A}z}^2\rangle}{\langle v_{\mathrm{A}}^2\rangle }\right)^{1/2} ,
\end{equation}
where $v_\mathrm{A}\propto B/\sqrt{\rho}$ is the Alfv\'en speed and $v_{\mathrm{A}z}$ its $z$-component. It is generally required that $Q_z\gtrsim 10$ for a satisfactory resolution of the vertical MRI modes. In our geometry we have $B_r \sim B_\theta$ and $B_\phi=0$ and thus $v_{\mathrm{A}z} \sim v_\mathrm{A}$. Therefore, with our choice $\beta=100$, the value of $Q_z$ ranges between $\sim$10 and $\sim$20 for our three resolutions, so all of them should yield satisfactory MRI evolution.

After the code initialization we let the torus evolve at given resolution. The state of the torus at $t_{\rm in}=25\,000\M$ is then taken as the initial state for the run with the moving star. Hence, the shape of the torus and the accretion rate profile just before perturbation also depend on the resolution. We compare the distribution of matter in the accretion disk by means of the averaged densities defined as 
\begin{eqnarray}
\left<\rho\right>_{(\theta,\phi)}(r) &=& \frac{\int^{2\pi}_0 \int^{\pi}_0 \rho \sqrt{-g} \,{\rm d}\theta {\rm d}\phi}{\int^{2\pi}_0 \int^{\pi}_0 \,\sqrt{-g}\, {\rm d}\theta {\rm d}\phi}, \\
\left<\rho\right>_{(r,\phi)}(\theta) &=& \frac{\int^{2\pi}_0 \int^{R_{\rm out}}_0 \rho \sqrt{-g}\, {\rm d}r {\rm d}\phi}{\int^{2\pi}_0 \int^{R_{\rm out}}_0 \sqrt{-g} \,{\rm d}r {\rm d}\phi}.
\end{eqnarray}

The profiles of these quantities at time $t_{\rm in}=25000\M$ are plotted in Fig.~\ref{Fig:rho} by dashed lines. While the angular shape of the torus is not affected by the resolution very much (except of the slightly different normalization), the radial shape of the flow exhibits various differences. The inner and outer edges of the tori approximately coincide but between them we observe differently located peaks and dips, which is also linked to slightly different accretion rates (see Fig.~\ref{Fig:mdot}). The onset of accretion and the initial rise of accretion rate is similar for each resolution, the same holds true for the character of the relaxation to a quasi-stationary level; however, higher resolution shows more variability. 

\begin{figure}
\begin{center}
\includegraphics[width=\linewidth]{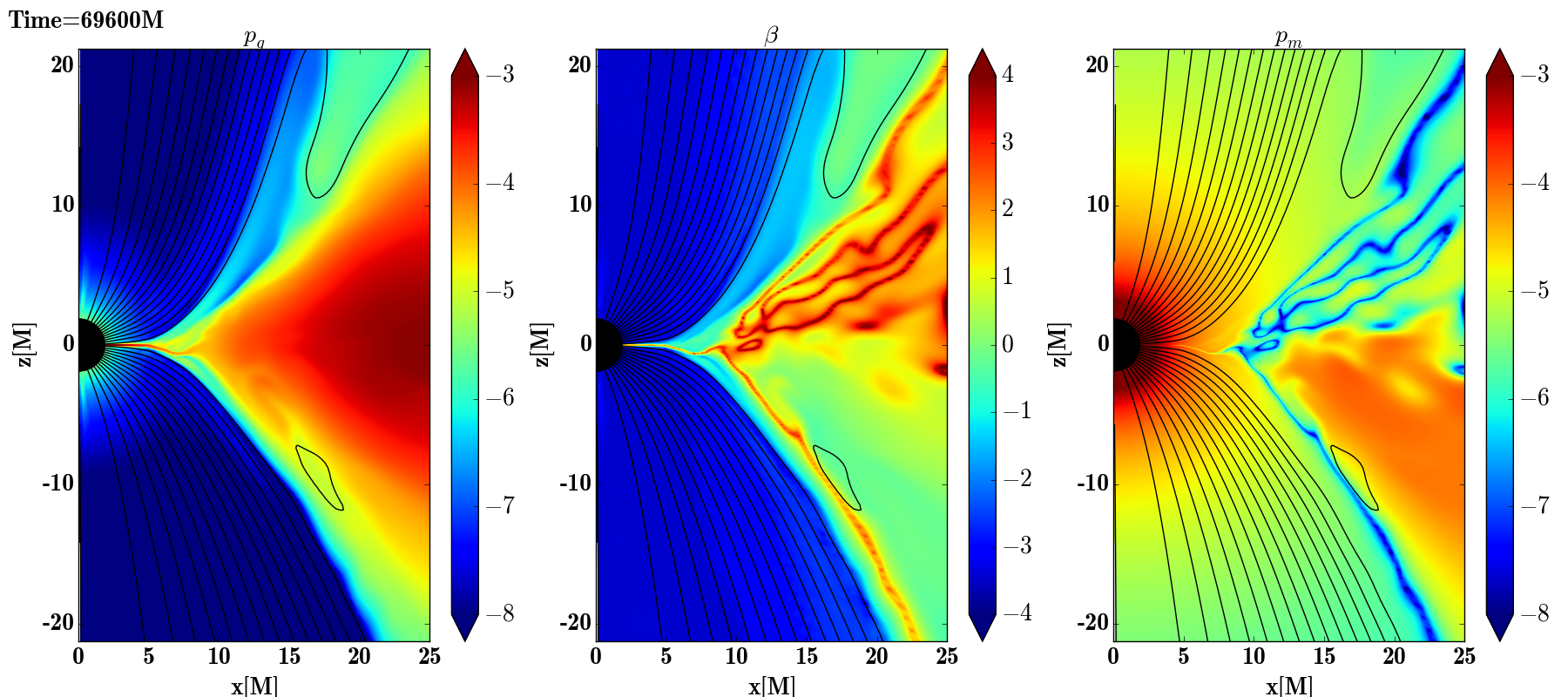}
\includegraphics[width=\linewidth]{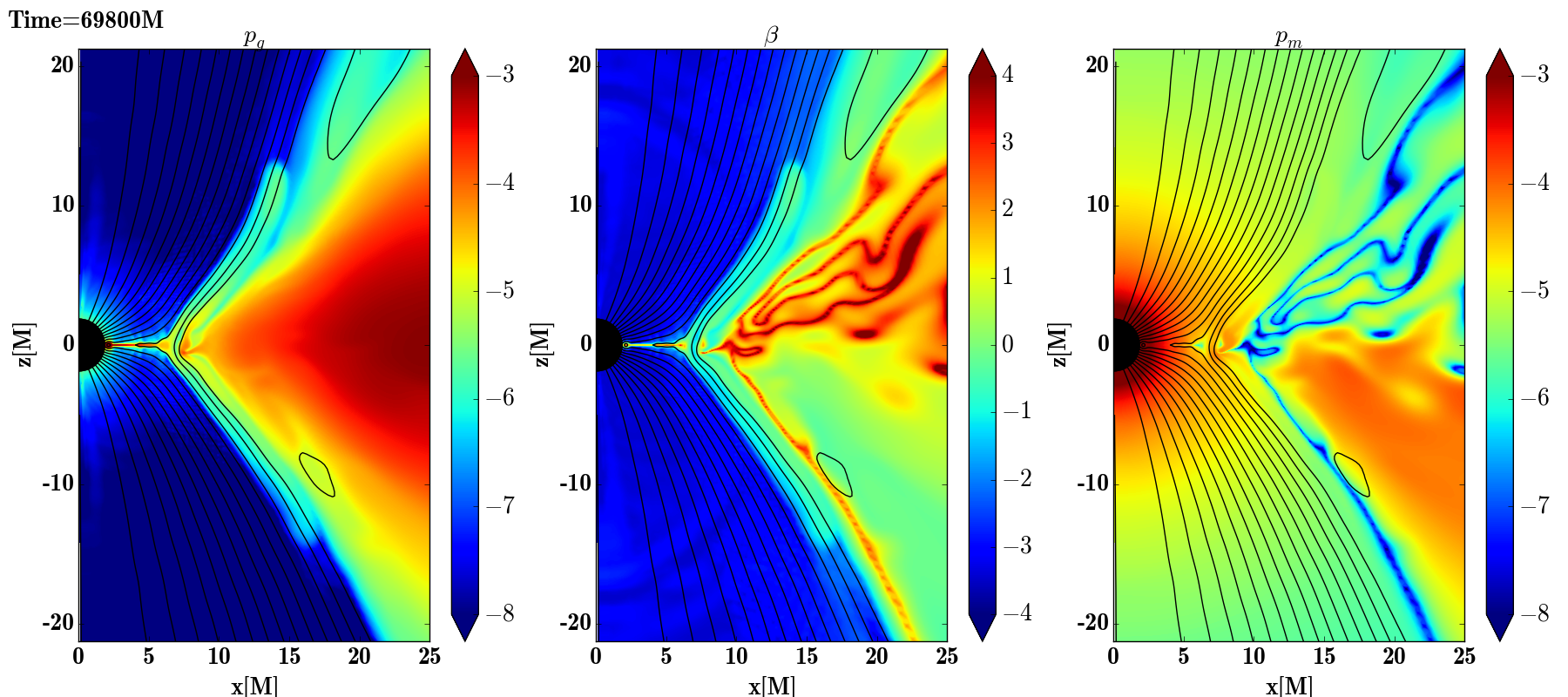}
\includegraphics[width=\linewidth]{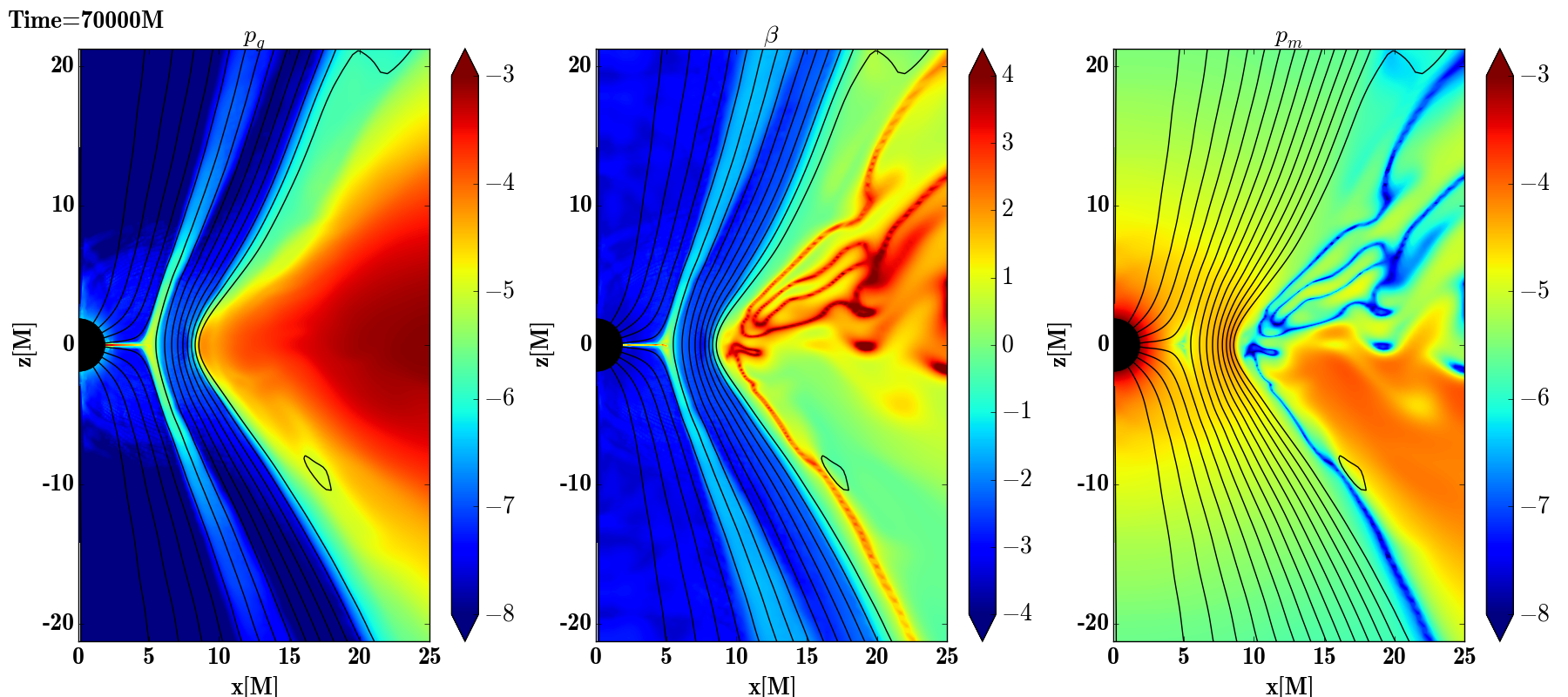}
\end{center}
\par\vspace{-1ex}\par
\caption{\label{Fig:V114-beta}Slices from run NP-HR at three time instances $t=69600{\M}, t=69600\M$ and $t=70000\M$. In the first column we show the gas pressure $p_g$, in the middle column the gas to magnetic pressure ratio $\beta$, and in the right column magnetic pressure $p_m$.   }
\end{figure}

\begin{table}[b]
    \centering
    \begin{tabular}{c|c|c|c|c|c|c}
         Run & $u_t$ & $u_\phi$ & $t_{\rm end} [{\rm M}]$ & $r [{\rm M}]$ & $\mathcal{R} [{\rm M}]$ & resolution  \\
         A & -0.9557 & 0.479 & $5\cdot10^4$ & 10 & 1&LR, FR, HR\\ 
         B & -0.9761 & 3.295 & $5\cdot10^4$ & 15 -- 25 & 1 &LR, FR, HR\\
         G & -0.9557 & 0.479 & $5\cdot10^4$ & 10 & 0.1& FR, HR\\ 
    \end{tabular}
    \caption{Summary of the star's orbital parameters in perturbed runs used to study the grid resolutions effects. $u_t$ and $u_\phi$ are the geodesic constants of motion in Kerr spacetime, $t_{\rm end}$ is the final time of the simulation, $r$ shows the radial range of the orbit, $\mathcal{R}$ is the radius of the star and in the last column we show with which resolution the case was computed.} 
    \label{Table:runs}
\end{table}
Most interestingly, the run NP-HR exhibits, for a certain time interval, a quasi-periodic flaring activity. This is due to the fact that during the evolution a dip in density in the torus forms when the inner part of the torus empties faster than new matter from the outer part of the torus comes inside. At a certain point, the density in the innermost region decreases until the flow is completely squeezed out of the equatorial plane and the magnetic field is reorganized. The reorganization happens at a radius of about $\sim10\M$, where the magnetic field lines transition into an ordered vertical field that stretches along the symmetry axis from the bottom to the top of the simulation domain while not intersecting the black hole (see the first and second row in Fig.~\ref{Fig:V114-beta}). At that point a small blob of matter is separated from the main body of the torus by the reconnected magnetic field lines and then it becomes quickly accreted into the black hole. The gas pressure at the inner edge of the torus is too low for the gas to be able to go through the strong magnetic field, hence the torus is detached from the black hole until new matter piles up and pushes the magnetic field lines back into the black hole. This repeats several times until the inner part of the torus fills again with enough gas to sustain stable accretion. Similar behaviour was seen also in some of the perturbed runs, where the inhibition of the matter inflow is caused by the motion of the star. 

\begin{figure}
\begin{center}
\includegraphics[width=\linewidth]{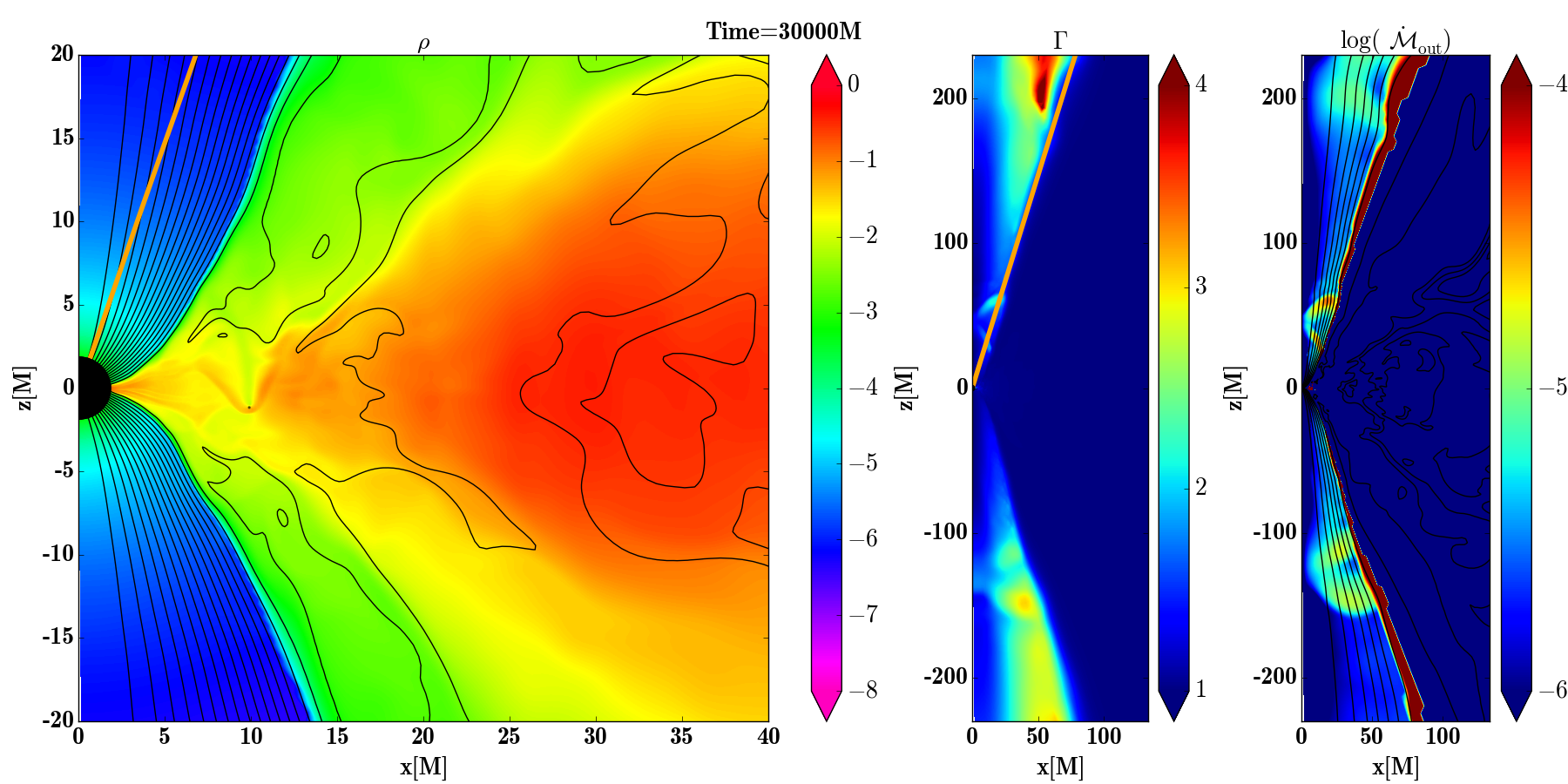}

\includegraphics[width=\linewidth]{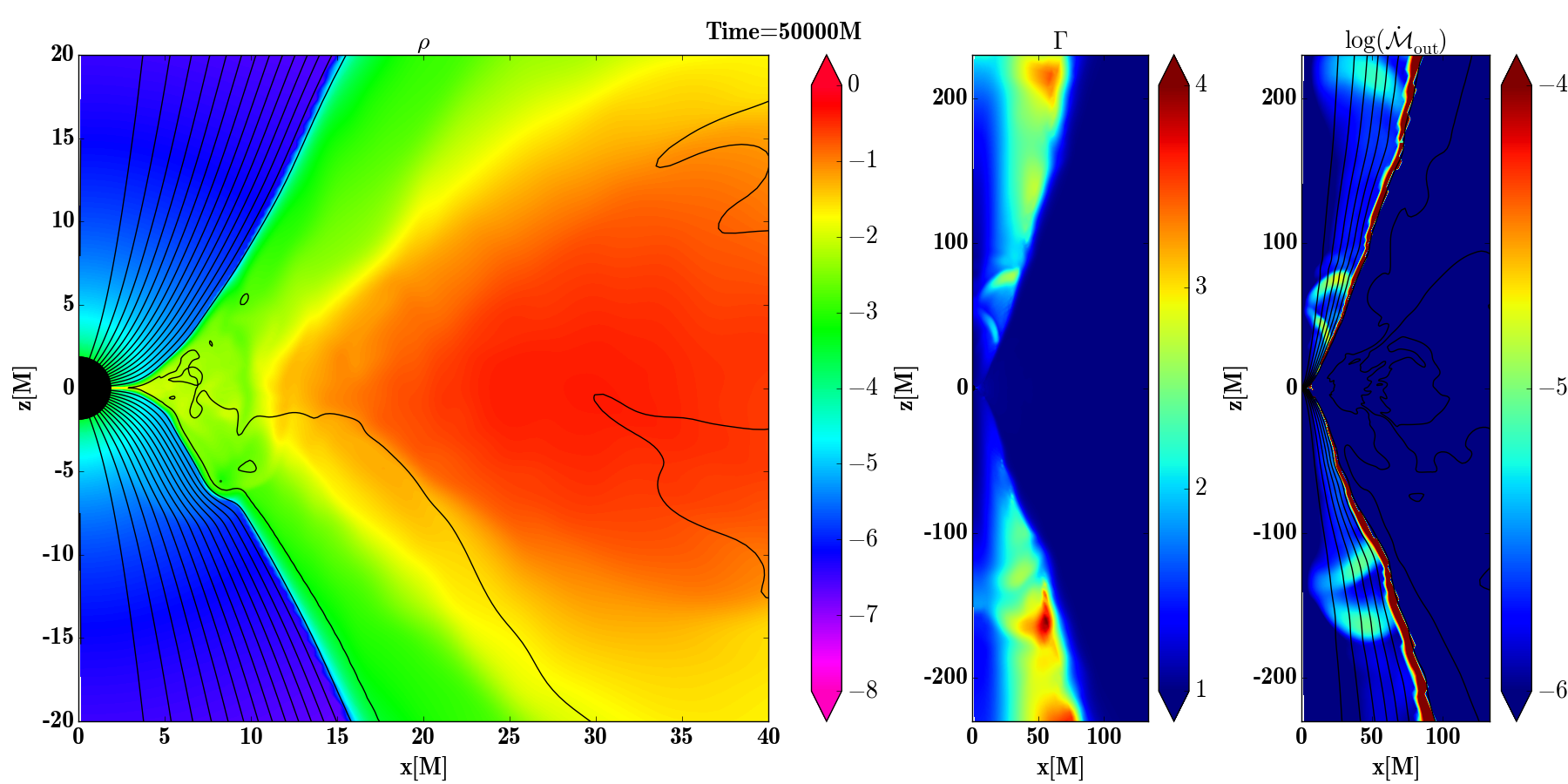}
\end{center}
\par\vspace{-1ex}\par
\caption{\label{Fig:V114-R1-slice}Slices from the simulation G-HR at time $t=30000\M$ (top) and $t=50000\M$ (bottom). The radius of the star $\mathcal{R}=0.1\M$ is on the lower limit constrained by the used resolution. In the top panel the opening angle (20°) of the funnel on the larger scale is denoted by orange straight line. The bow shock caused by the moving star can be seen expanding into the black hole and also outwards into the torus. The star expels gaseous blobs into the funnel. In the bottom panel the reduction of density of accreting gas below the star orbit is visible. The blobs are still outflowing in a quasi-periodic manner.}
\end{figure}

\subsubsection{Interaction of the star with the flow}
We chose three different cases of the perturbing star and ran the simulations with LR, FR and HR. These runs are denoted as A,B and G (in accordance with (Sukov\'a et al., 2021, work in progress)) and the orbital parameters of the star motion for each run are summarised in Table~\ref{Table:runs}. Orbit~A corresponds to a nearly circular orbit going close to the black hole rotational axis, orbit~B is then embedded in the accretion torus. The radius of the perturbing star is $\mathcal{R}=1\M$.  Orbit~G is the same as orbit~A, but the radius of the star is smaller, $\mathcal{R}=0.1\M$. All the choices of star orbits induced pronounced effects on the structure of the accreting torus as well as on the time dependence of the accretion rate.

\begin{figure}
\begin{center}
\includegraphics[width=\linewidth]{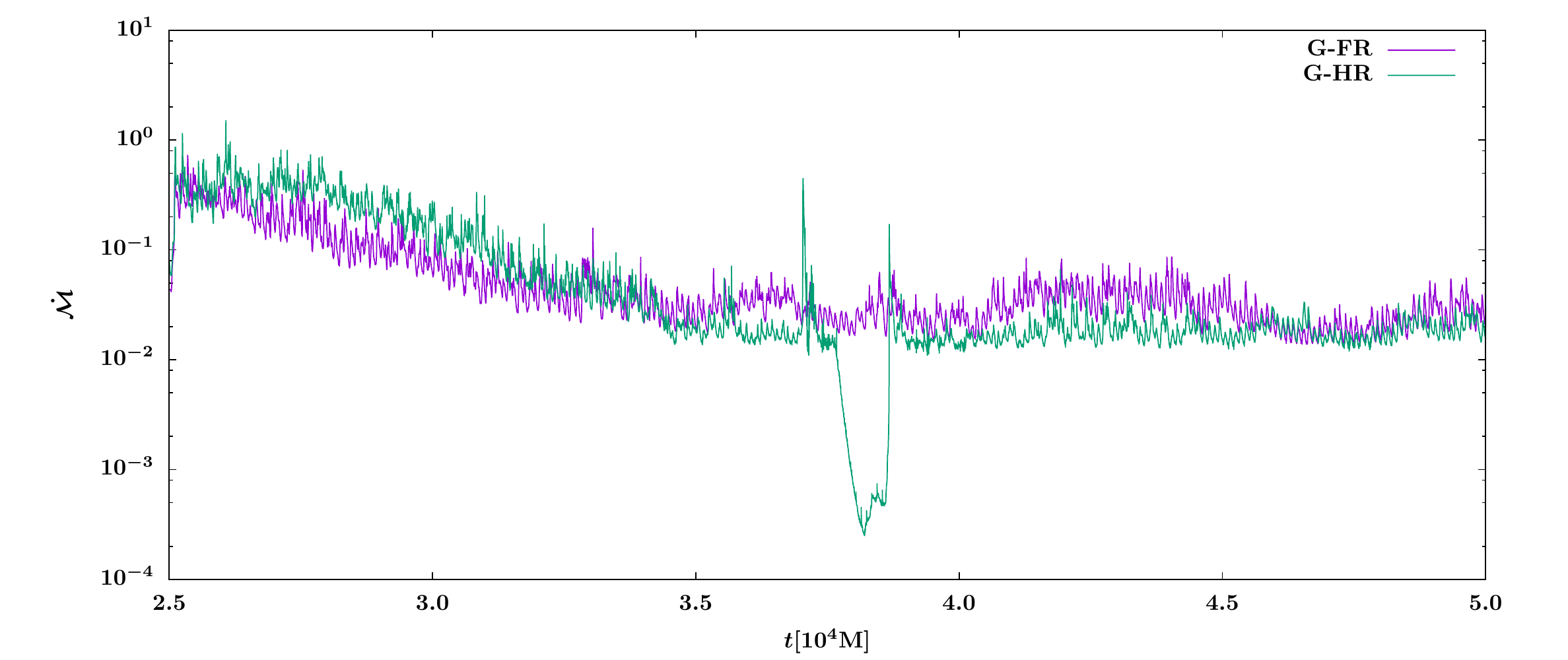}

\includegraphics[width=\linewidth]{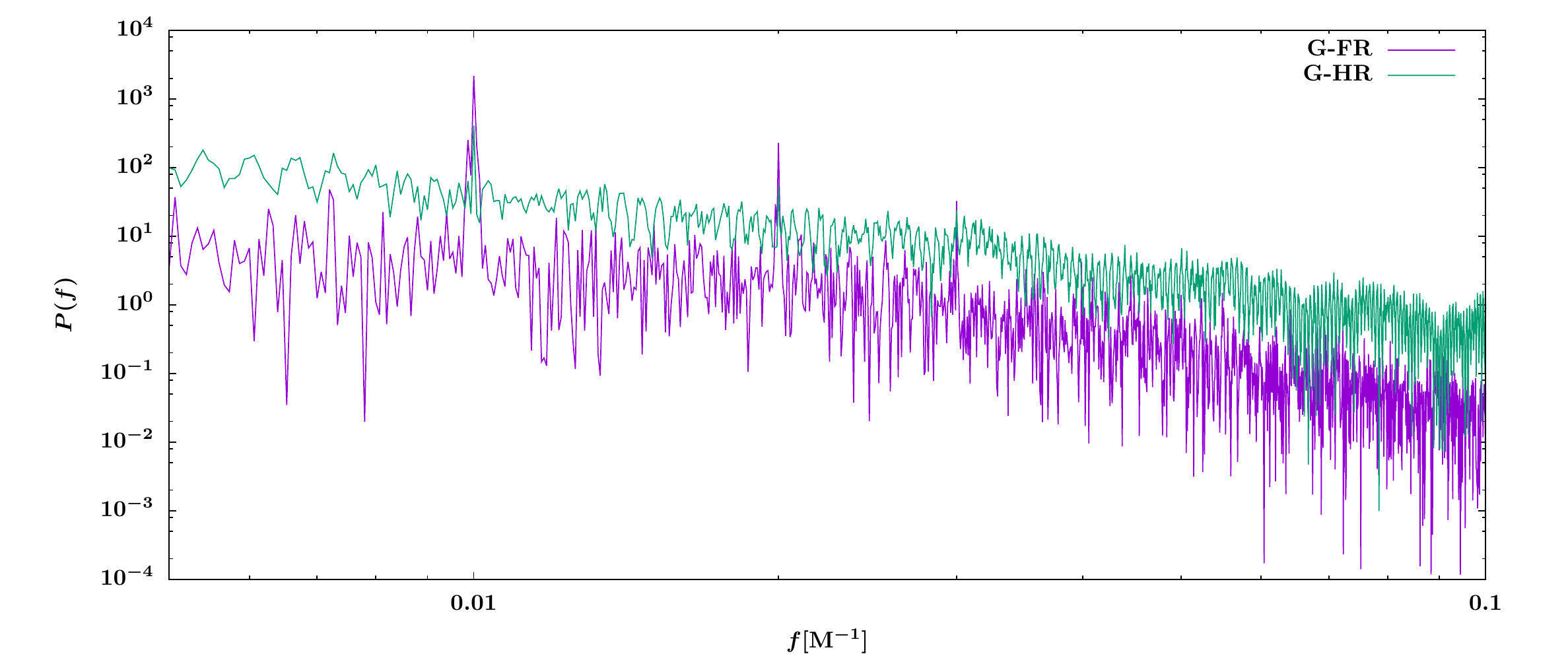}
\end{center}
\par\vspace{-1ex}\par
\caption{\label{Fig:mdot-runG}Time dependence of the accretion rate $\dot{\mathcal{M}}$ for runs G-FR and G-HR. The corresponding PSD computed from the settled state $t\in(3.5,5)\cdot 10^4\M$ (bottom). }
\end{figure}

We present slices from the simulation runs which contain three plots. On the first plot on the left the density $\rho$ in (arbitrary) code units in logarithmic scale is shown, in the middle plot we show the Lorentz factor $\Gamma$ of the gas and in the third plot the ``outflowing accretion rate'' is given, as computed according to 
\begin{eqnarray}
    \dot{\mathcal{M}}_{\rm out} &=& \rho \, \frac{u^r}{u^t} \sqrt{-g} \, {\rm d}\theta \, {\rm d}\phi \qquad {\rm for \, } \Gamma > 1.155, \\
    \dot{\mathcal{M}}_{\rm out} &=& 0 \hspace{2.85cm} {\rm for \, } \Gamma \leq 1.155\,.
\end{eqnarray}
Such definition ensures that we follow only the rapidly outflowing gas in the funnel region and not the slowly moving gas in the torus.
The second and third plots display a larger portion of the computational grid, however this is still zoom into the inner part of the grid, which spans up to $r_{\rm out}=2\cdot10^4{\rm\,M}$. The example is given in Fig.~\ref{Fig:V114-R1-slice}, where the HR version of run~G at time $t=30000\M$ and at the end of the run $t_f=50000\M$ is shown. 

The HR simulation of run G exhibits qualitatively similar results as the FR run, including the choking of the torus in the innermost part, the consequent decrease of accretion rate by approximately one order of magnitude, and the existence of blobs of matter outflowing mainly along the boundary between the funnel and the torus in a quasi-period manner. The accretion rate and the power spectrum density (PSD) obtained by the Fourier transform of the accretion rate are plotted in the top and the bottom panels of Fig.~\ref{Fig:mdot-runG}, respectively. Even though the FR resolution is barely capturing the star in the grid, the mean value of the accretion rate decreases in a similar manner in the FR and HR runs. The HR run, however, exhibits a few larger peaks and one significant drop of accretion rate, while in the FR run such substantial variability is not seen. In both cases, the PSD shows a pronounced peak at $f=10^{-2}\M^{-1}$, which corresponds to half of the orbital period of the star. We could not repeat this run with LR, since it has a grid too sparse to capture such a small star.

\begin{figure}
\begin{center}
\includegraphics[width=\linewidth]{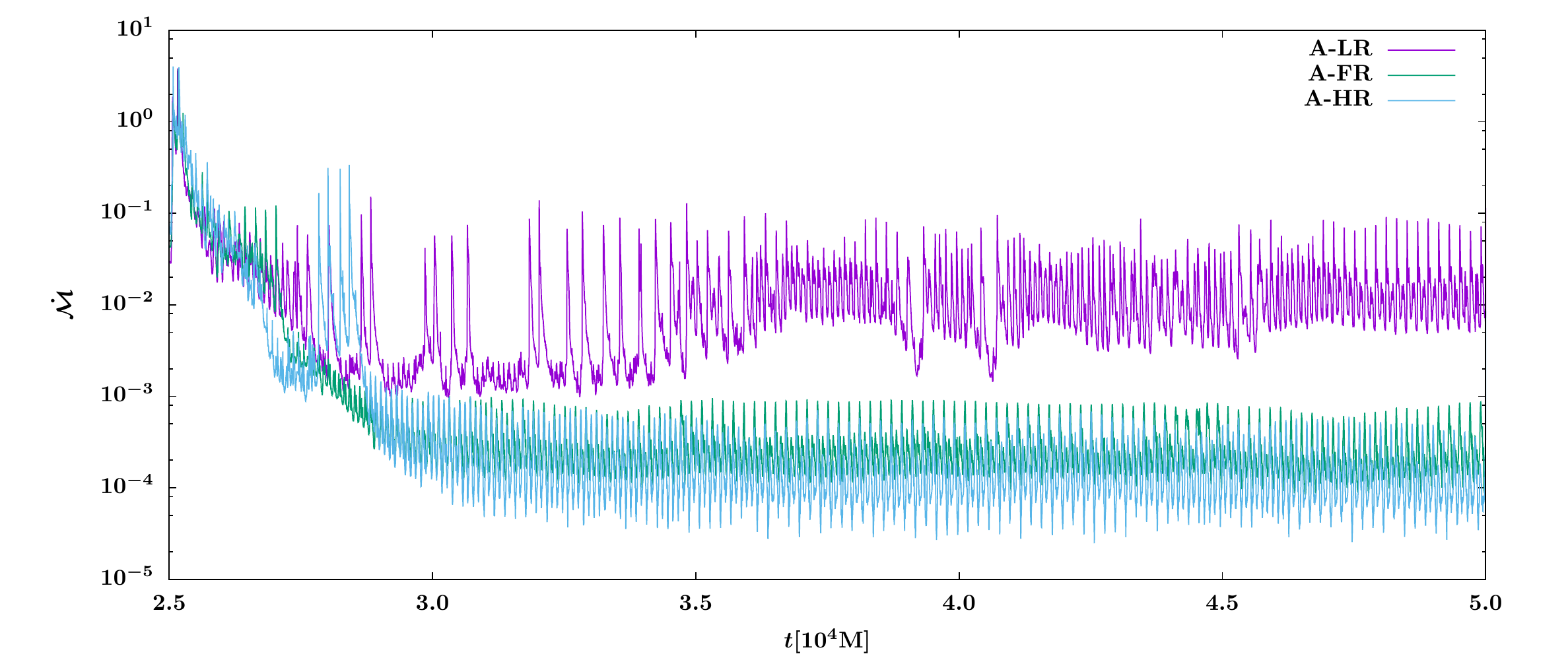}

\includegraphics[width=\linewidth]{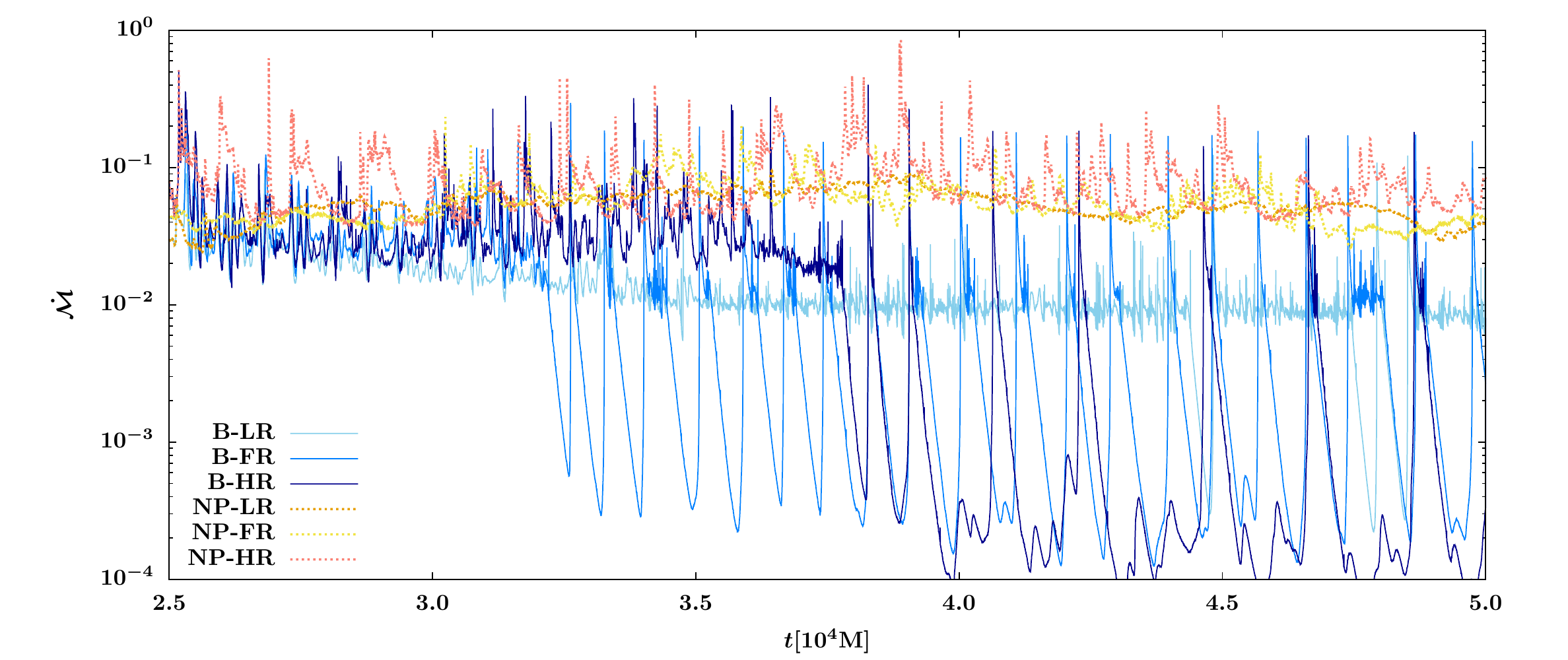}
\end{center}
\par\vspace{-1ex}\par
\caption{\label{Fig:mdot-runA}Time dependence of the accretion rate $\dot{\mathcal{M}}$ for the runs perturbed by star A with LR (purple), FR (green) and HR (blue) in the top panel. Bottom panel shows the same for runs with star B (solid lines) and non-perturbed runs (dotted lines).}
\end{figure}

For runs with star A the accretion rate profile for all the three resolutions is given in Fig~\ref{Fig:mdot-runA}. The initial decrease of the accretion rate is similar for each case, however, the LR run exhibits fewer periodic peaks with a higher amplitude and it settles at a larger accretion rate than the FR and HR runs. The two runs with higher resolution coincide very well except of a transient flaring period and a slightly lower mean accretion rate value in the HR case. The radial and angular distributions of the gas at the end of the run is shown in Fig.~\ref{Fig:rho} with solid lines. While the LR run shows higher peaks and a dip in the center, the shapes of FR and HR profiles coincide quite well. The quasi-periodic features in the accretion rates are observable for runs with all resolutions and the power spectra of runs A-FR and A-HR coincide very well, while in A-LR case the corresponding peaks are weaker.

The FR resolution of run B was found in a substantially flaring state with variations in the accretion rate spanning more than three orders of magnitude. The computation with LR shows a similar decrease of the accretion rate as FR, however, only three episodes of big dips and peaks occurred in this case. The amplitude of the oscillation is almost the same, while the time duration of the dips is a little shorter than in FR. Interestingly, in the HR case it also takes a longer time to develop the flaring state and the dips are longer, whereas the peaks have a similar duration.

\subsection{Effect of the used approximation to the star motion}

\begin{table}[b]
    \centering
    
    \begin{tabular}{c|c|c|c|c|c}
         Run & IA-pi/64 & IA-pi/4 & A-front face & A & A-one way\\
         $\bar{\mathcal{M}}$ & $3.2\cdot 10^{-2} $ &$7.8\cdot 10^{-2}$  & $2.9\cdot 10^{-4}$ & $2.8\cdot 10^{-4}$ & $1.3\cdot 10^{-4}$  \\
         $\mathcal{A}$ & 22.5 & 6.57 & 2.78 & 3.17 & 5.37 \\
         $f_1[\M^{-1}$]&0.0049&0.0049&0.010&0.010&0.005\\   
         $f_2[\M^{-1}$]&0.0148&0.0099&0.005&0.005&0.010\\
         $f_3[\M^{-1}$]&0.0099&0.0149&0.020&0.020&0.015\\
         
    \end{tabular}
    \caption{The mean accretion rate $\bar{\mathcal{M}}$ and the amplitude of oscillations $\mathcal{A}$ during the settled state $t\in(35000,50000)\M$ of the runs with different star realisations, for which $t_{\rm end} = 5\cdot10^4\M$, $r=1{\rm M}$, $\mathcal{R} = 1\M$ with resolution FR.  We have 2 runs with the impulse approximation, IA-pi/64 perturbs cells in an annulus sector with the angular width $\Delta \theta = \pi/64$ and radial width $\mathcal{R}$, while in IA-pi/4 the annular sector of the perturbed region is much larger ($\Delta \theta = \pi/4$). In Run A-front face we perturb cells in the disc shaped region with $\Delta \theta = \pi/4$ moving along the geodesics, in run A the star is described as a sphere with radius $\mathcal{R}$. Run A-one way differs from run A in the way that the perturbation is turned on only when the star moves "downwards", that is $v^\theta= \frac{{\rm d} x^\theta}{{\rm d} t}>0$.  The frequencies of the first three highest peaks in the power spectrum $f_1, f_2, f_3$ is in the last three rows.  \label{Table:runs-realisation}}
    
\end{table}

The effect of a solid star equipped with a magnetic field and a stellar wind moving in the accretion flow is very complicated. We have to simplify the picture to be able to describe such a process in our simulations. We focus on the dynamical effect of the star on the accreting gas, and we thus neglect the possible feedback of the gas on the star motion or structure. Therefore, the star can be treated as a test body moving along a geodesic trajectory. We also neglect the possible accretion of the gas on the star (even if the star was to be understood as a stellar-mass black hole) or strong wind outflow from the star that could enrich the accretion flow. Therefore, we consider the star to be only a solid body, which is pushing the gas along its trajectory. However, still several different simple approximations of this scenario can be used. Here we compare results from the so-called impulse approximation and a gradual perturbation of the gas by the moving star. We use the fiducial resolution  $n_r=252, n_\theta=192$.

\subsubsection{Impulse approximation}\label{Sec:Impulse-app}
The most simplistic approach is the so-called impulse approximation (IA), where the transit of the star is simulated such that the gas in the tube corresponding to the volume through which the star moves gets the impulse by the star at the moment when the star passes through the equatorial plane \citep{1991MNRAS.250..505S,1993MNRAS.265..365V}. This is a particularly well substantiated description for supersonic transits. We consider a star moving on a circular orbit, compute its orbital frequency and with this frequency we periodically set the velocity of the gas as equal to the orbital velocity of the star. The perturbed region is thus described by the following relations,
\begin{eqnarray}
    |(r-r_{\rm star})| &<& \mathcal{R}, \\
    |(\theta-\theta_{\rm star})| &<& \Delta \theta,\\
    t_{\rm perturb}&=&n T_{\rm orb}.
\end{eqnarray}

The free parameter of this approximation is the angular width of the perturbed region, which we set to two different values: $\Delta \theta = \pi/4$ in run IA-pi/4 and $\Delta \theta = \pi/64$ in run IA-pi/64. In this way we either perturb almost all gas of the torus along the path of the star in the former case or only a relatively small region close to the equatorial plane in the latter case.

\subsubsection{Gradual perturbation}\label{Sec:Geodesic-star}
The second approach considers a sequential action of the moving star on the gas - we called it gradual perturbation (GP). We solve the geodesic equation for the star motion along with the GRMHD evolution of the plasma. Then in each time step we change the velocity of the gas as equal to the velocity of the star within a region ascribed to the star. In this way we simulate the fact that the solid body of the star moves in the grid at a given velocity. 

To study also the effect of the exact shape of the moving body, we set the region corresponding to the star in two different ways:

\begin{enumerate}
    \item We consider a disc-shaped region with the radius equal to the star radius, which is only several zones wide in the $\theta$-direction. Hence, the perturbation is done in the domain consisting of cells satisfying
    \begin{eqnarray}
        |(r-r_{\rm star})| &<& \mathcal{R}, \\
        |(\theta-\theta_{\rm star})| &<& \Delta \theta \qquad \Delta \theta = 4 \pi/n_\theta = \pi / 48,
    \end{eqnarray}
    where $r,\theta$ are BL coordinates of the grid cell center. 
    This corresponds to a "front face" of the star moving in the flow and is used in run A-front face.
    
    \item All grid cells with cell centers located at ($r,\theta$) in BL coordinates occupying the full volume of the star described in 2D simulations by
    \begin{eqnarray}
        {(\Delta x)}^t &=& 0,\\
        {(\Delta x)}^r &=& r-r_{\rm star},\\
        {(\Delta x)}^\theta &=& \theta-\theta_{\rm star},\\
        {(\Delta x)}^\phi &=& 0,\\
        d &=& \sqrt{ g^{\rm BL}_{\alpha \beta}\Delta x^{\alpha}\Delta x^{\beta}},\\
        d&<&\mathcal{R}
    \end{eqnarray}
    are perturbed. This approach, which we consider as the best available description of the star, will be used in Sukov\'a et al. (2021, work in progress) for all simulations. 
\end{enumerate}

\begin{figure}
\begin{center}
\includegraphics[width=\linewidth]{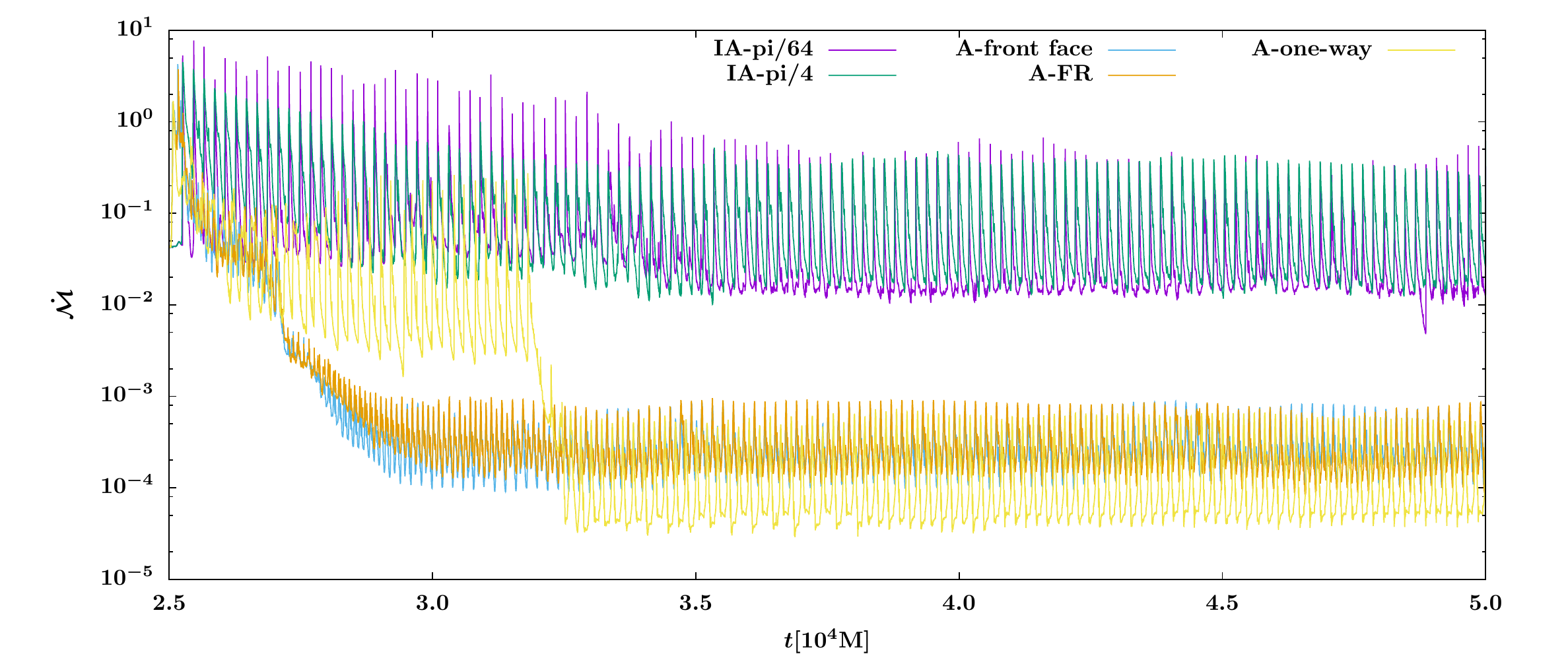}

\includegraphics[width=\linewidth]{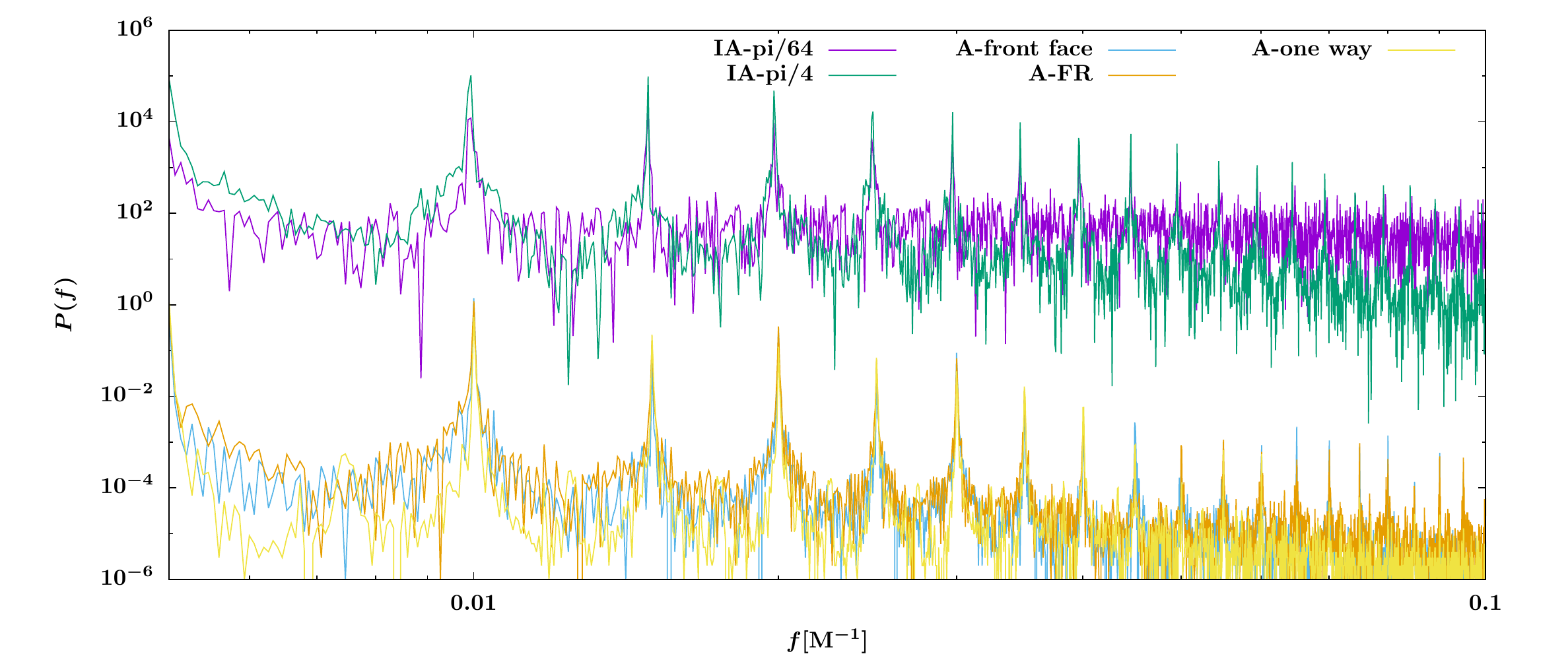}
\end{center}
\par\vspace{-1ex}\par
\caption{\label{Fig:mdot-approx}Time dependence of the accretion rate $\dot{\mathcal{M}}$ for runs with different star realizations (top). The corresponding power spectrum is computed from the settled state $t\in(35000,50000)\M$ (bottom). }
\end{figure}

\vspace{0.1cm}
The comparison of the aforementioned approaches is done on the example of (nearly) circular orbit passing perpendicularly through the equatorial plane with $r_{\rm star}=10\M$, which was  with the last realization of the star already shown as run A-FR in Sec.~\ref{Sec:resolution}. All the runs are initiated with the evolved torus at $t_{\rm in}=25000\M$ similarly as our other computations and the fiducial resolution is used.
The accretion rates for all types of star realizations are compared in the top panel of Fig.~\ref{Fig:mdot-approx}, while the PSD is shown on the bottom panel. 

We compare the accretion rate profile, its mean value $\bar{\mathcal{M}}$ and the amplitude of the peaks computed according to 
\begin{equation}
\mathcal{A} = {\rm abs} \left( \frac{ \max(\mathcal{M}) - \min(\mathcal{M})}{\bar{\mathcal{M}}} \right),
\end{equation}
where the average, maximal and minimal values are taken from the settled time period $t\in(35000,50000)\M$. The results are summarised in table~\ref{Table:runs-realisation}.

The most prominent difference between the two approaches is seen in the case of IA used in runs IA-pi/4 and IA-pi/64: the accretion rate decreases only by about one order of magnitude, while in the case of GP, the accretion rate drops by more than three orders of magnitude. The temporal profile of the accretion rate shows quasi-periodical structures. While in all runs, we can clearly see the presence of the orbital frequency of the star, which is $f_{\rm orb}=4.955\cdot10^{-3}\M^{-1}$, and its multiples, the peaks of the accretion rate are larger in case of impulse approximation. 

In contrast to the used approximation, the exact realization of the perturbed region does not affect the results significantly. The accretion rate mean values as well as the frequency and amplitudes of the oscillations coincide very well for the pairs of runs IA-pi/64 and IA-pi/4 and A-front face and A. Therefore, the evolution of the gas depends only weakly on the exact shape of the star, as long as the radius of star remains the same.

\subsubsection{One-way transit of the star}
Two-dimensional simulations are simplified and incomplete due to the imposed  ``squeezing'' $\phi$ (azimuthal) direction into a single 2D slice. This averaging can lead to some artificial effects in the gas evolution because in reality the star transits through the disc in one direction at one half of the disc and in the other direction in the opposite half of the disc, while in our simulations the star passes through the disc in both directions at similar place.  Therefore, we have repeated run A with the complete (3D) geodesic motion, so that the perturbation is turned on only when $v^\theta= \frac{{\rm d} x^\theta}{{\rm d} t}>0$ (that is when the star moves ``downwards'' in our slice), which reflects more accurately the local evolution of the gas (run A -- one way). 

The corresponding accretion rate and its power spectrum are shown in Fig.~\ref{Fig:mdot-approx}. We can see that the transient time at the beginning of the simulation lasts longer with quite high peaks and higher values of the accretion rate, however after $\sim8000 M$ the accretion rate drops to even slightly smaller values than in run A. The power spectrum shows that in case of run A -- one way, the most prominent peak with $f=5\cdot 10^{-3}\M^{-1}$ corresponds to the star orbital frequency, while in case of run A and A -- front face the peak at its double value $f=10^{-2}\M^{-1}$ gets more power, which is in accordance with our expectations.

\begin{figure}
\begin{center}
\includegraphics[width=\linewidth]{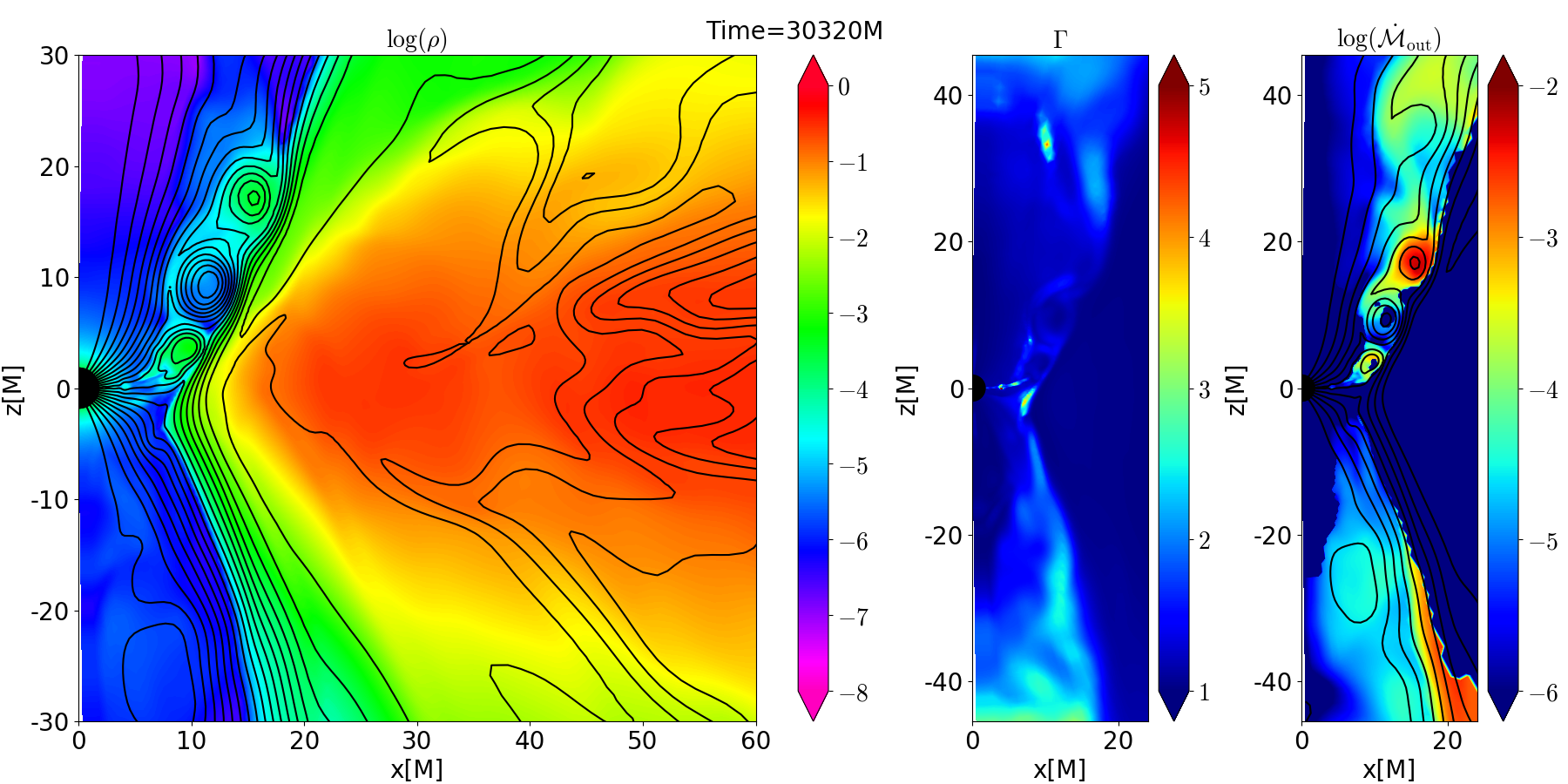}

\includegraphics[width=\linewidth]{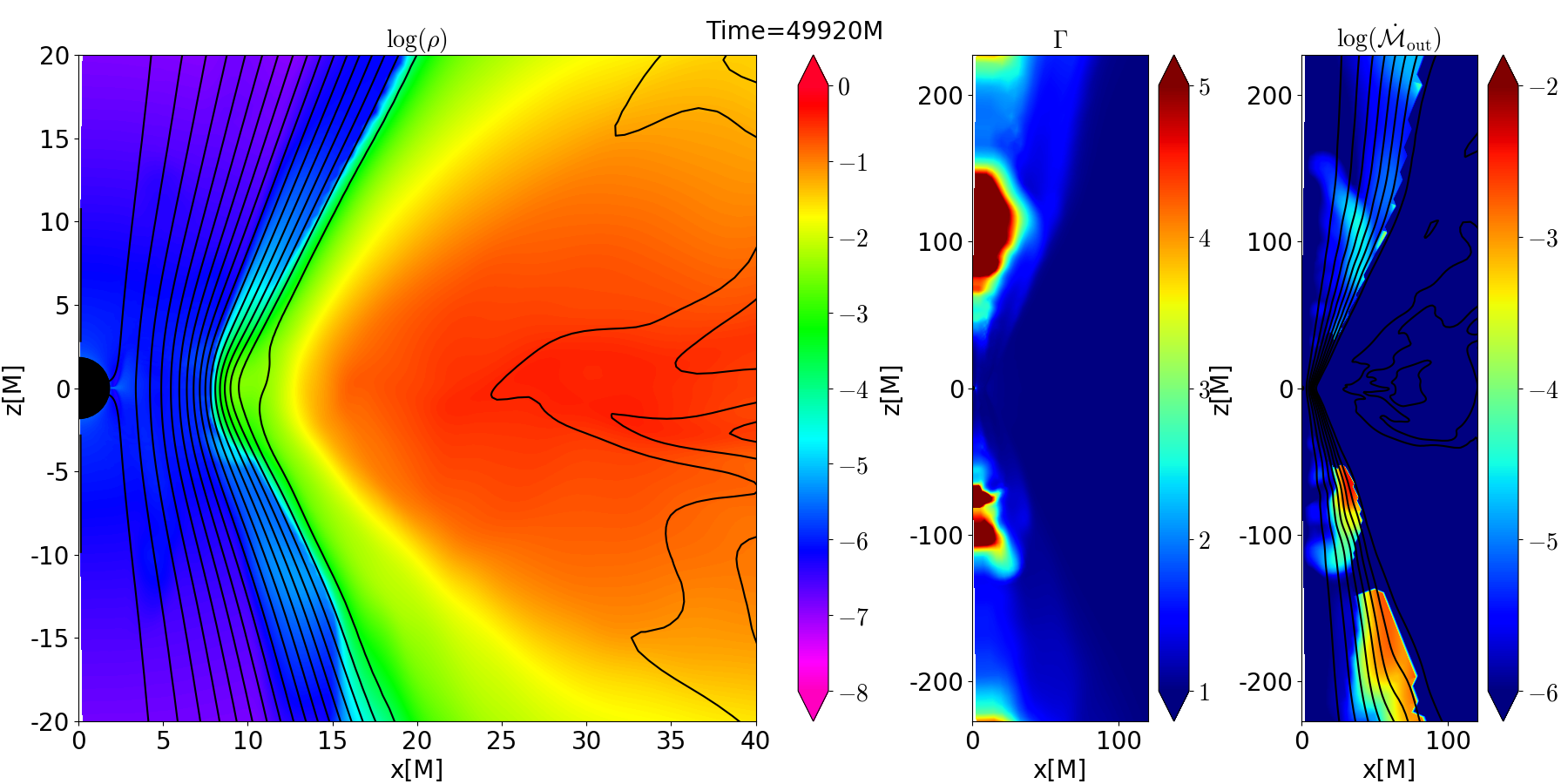}
\end{center}
\par\vspace{-1ex}\par
\caption{\label{Fig:V113-R42-slice}Slices from the run A-one way. The top panel shows the time $t=30320\M$ before the accretion rate drops down. The blobs encircled by magnetic field lines outgoing upwards along the boundary between the funnel and the torus can be seen. On the bottom panel the situation at $t=49920\M$ illustrates the detaching of the torus from the black hole, the existence of long vertical magnetic field lines, which do not end in the black hole, and large blobs of gas outflowing downwards, while small blobs are outflowing upwards.}
\end{figure}

The evolution of the gas shows one interesting feature distinct from the run A, which is the formation of blobs wrapped in magnetic field-line loops that develop and depart toward the opposite direction than the motion of the star (i.e. ``upwards'' -- see top panel of Fig.~\ref{Fig:V113-R42-slice}). Later, when the accretion rate decreases, the outflowing blobs of matter proceed asymmetrically with respect to the equatorial plane. There are larger blobs going downwards along the funnel boundary, but there are also fainter blobs moving upwards -- see the bottom panel of Fig.~\ref{Fig:V113-R42-slice}. Because in reality the star moves in the opposite direction through the other half of the disc, there will be an asymmetric outflow of matter also with respect to the rotational axis and the position of stronger and fainter outflow will rotate around the axis with the precession frequency of the star orbit. The possible helical trajectory of the expelled gas has to be studied in 3D simulations.

\section{Discussions and Conclusions}\label{conclus}
In the present paper we examined the effect of repetitive transits of a model star across the accretion slab. We focused on the role of the numerical set-up of the adopted scheme, the effects of grid resolution and the approximation used for the passages of the star. Even the current simplified approach indicates an interesting possibility of influencing the accretion rate by the repetitive transits and ejecting plasmoids from the inner disk with quasi-periodic signatures of the stellar orbit.

The comparison of LR, FR and HR runs has shown that the FR and HR yield very similar results, both qualitatively and quantitatively, while the LR runs, which probably render the MRI poorly, differ more significantly. However, the flaring state of the torus is still quite sensitive to the resolution of the grid. We can attribute this to the fact that, in the flaring state, the inner part of the torus is squeezed to a very thin layer, hence the description of the exact process of the blob separation and the following accretion needs a very high resolution along the equatorial plane, both in the $r$ and $\theta$ directions. It is then left for a future study with a better resolution of the inner region to capture the shape, amplitude and frequency of the peaks in the flaring state more accurately.

The runs with different orbits showed that the presence of the star has a substantial effect on the accretion flow for various configurations. We  observed that the accretion can be effectively inhibited, in particular when the star moves close to the black hole (on the radii $\sim 10 {\rm\,M}$), where a drop of the accretion rate by three orders of magnitude was found. We also observed blobs of matter expelled from the torus into the empty magnetised funnel region. The blobs are then magnetically accelerated outwards with mildly relativistic speeds along the boundary between the funnel and the torus, which has an opening angle of about 20$^{\circ}$ with respect to the rotation axis. These results will be studied in a separate study (Sukov\'a et al., 2021, work in progress).

We expect that the main features found in our simulations, such as the decrease of accretion rate, presence of quasi-periodic features stimulated by the orbital frequency of the star and outflowing blobs, are described sufficiently well by our FR and HR resolution, so these results should persist in more detailed computations. The simulation with one-way transit of the star has shown that in the full 3D case we can expect a non-axisymmetrical ejection of plasmoids in the funnel region. They can represent a spot rotating around the axis with the precession frequency of the star orbit.

Let us note that our computations were performed while assuming negligible radiative cooling of the flow. Hence our results are applicable mainly in the low-luminous galactic nuclei, such as Sgr A* in the center of our galaxy
\citep{2014ARA&A..52..529Y}.  Even though the simulations with radiative cooling have shown that even in case of Sgr~A* for some observationally allowed values of accretion rate the inclusion of cooling has an effect on the accretion torus \citep{Sasha-radiative-cooling}, the overall structure of the accretion flow remains similar as in the non-cooled state. If the accretion rate becomes higher, reaching about three to one order bellow the Eddington accretion rate, the cooling becomes substantial and the flow transforms into a cold Keplerian accretion disc. The issue of stellar transits in active galactic nuclei with high accretion rates will, therefore, require the inclusion of cooling into the scheme. 



\ack

The authors acknowledge the Czech Science Foundation - Deutsche Forschungsgemeinschaft collaboration project (GAČR 19-01137J), and the Czech-Polish mobility program (M\v{S}MT 8J20PL037). MZ acknowledges the financial support by the National Science Center, Poland, grant No. 2017/26/A/ST9/00756 (Maestro 9) as well as the NAWA financial support under the agreement PPN/WYM/2019/1/00064 to perform a three-month exchange stay at the Charles University and the Astronomical Institute of the Czech Academy of Sciences in Prague. VW was supported by European Union’s Horizon 2020 research and innovation programme under grant agreement No 894881.


\bibliographystyle{ragtime}
\bibliography{suk}

\end{document}